\documentclass[twocolumn,noshowpacs,preprintnumbers,amsmath,amssymb,10pt]{revtex4-1}
\usepackage{caption}
\usepackage[justification=centerlast]{caption}
\usepackage{graphicx}
\usepackage{subfigure}
\usepackage{dcolumn}
\usepackage{bm}
\usepackage{subeqnarray}
\usepackage{amsmath}
\usepackage{color}
\usepackage{esvect}
\usepackage{ulem}

\usepackage{verbatim}
\usepackage{ragged2e}
\usepackage{braket}

\newcommand{\LN}{LiNbO$_3$ }
\newcommand{\Fig}[1]{Fig.~\ref{#1}}

\newcommand{\degC}{$^o$C}

\newcommand{\Mod}[1]{\textcolor{black}{#1}}
\newcommand{\ModTo}[1]{\textcolor{black}{#1}}

\emergencystretch=\maxdimen
\begin{document}

\title{Nonlinear integrated quantum electro-optic circuits}

\author{Kai-Hong Luo$^{\star}$, Sebastian Brauner, Christof Eigner, Polina R. Sharapova,\\
 Raimund Ricken, Torsten Meier, Harald Herrmann, Christine Silberhorn}

\affiliation{Department of Physics and CeOPP, University of Paderborn,
Warburger Str.~100, 33098, Paderborn, Germany}

\begin{abstract}
Future quantum computation and networks require scalable \Mod{monolithic} circuits, which incorporate various advanced functionalities on a single physical substrate. \Mod{Although substantial progress for various applications has} already been demonstrated on different platforms, the range of diversified manipulation of photonic states on demand \Mod{on a single chip has} remained limited, especially dynamic time management. Here, we demonstrate an electro-optic device, including photon pair generation, propagation, electro-optical path routing, as well as a voltage-controllable time delay of up to $\sim$12 ps on a single Ti:\LN waveguide chip.
As an example, we demonstrate Hong-Ou-Mandel interference with a visibility of more than 93$\pm$ 1.8\%. Our chip \Mod{not only enables} the deliberate manipulation of photonic states by rotating \Mod{the} polarization but also provides precise time control. Our experiment reveals that we have full flexible control over single-qubit operations by harnessing the complete potential of fast \Mod{on-chip electro-optic modulation}.
\end{abstract}

\maketitle

\section*{Introduction}

For \Mod{the} future deployment of practical quantum communication and information systems, advanced integrated quantum devices should comprise several sections: quantum state generation, path, power, and/or polarization routing, as well as phase or polarization manipulation, temporal and spectral synchronization, and ultimately also detection.
In the past decade, many optical circuits for quantum gates \cite{PolitiScience2008,CarolanScience2015}, quantum interference \cite{SilverstoneNP2014}, quantum metrology \cite{GiovannettiNP2011}, boson sampling \cite{TillmannNP2013,BroomeScience2013,SpringScience2013}, and quantum walks \cite{PeruzzoScience2010,CrespiNP2013} in different materials have been demonstrated. Most of these circuits are realized in \Mod{$\chi^{(3)}$ materials such as glass \cite{SpringOptica2017}, silicon nitride \cite{MossNP2013}, silicon-on-insulator \cite{SilverstoneIEEE2016}, and silica-on-silicon \cite{Faruque1705,SantagatiJO2017,MatsudaOE2014}}. \Mod{By contrast, the development} of integrated photonic devices based on second-order nonlinearities \cite{JinPRL2014,VergyrisSR2016,AlibartJO2016,LenziniLPR2017} has fallen far behind, despite \Mod{the fact that} exploiting $\chi^{(2)}$ nonlinearities is much more efficient. In particular, the full potential of fast active electro-optic routing and rotation of polarized photons in integrated quantum circuits has not yet been \Mod{fully} harnessed, despite the successes with tunable couplers \cite{MartinNJP2012} and voltage-controlled phase shifters \cite{JinPRL2014,Midolo1707}. \Mod{Although} substantial progress toward actual applications has been achieved, \Mod{to date, there exists no quantum electro-optic device, which can generate and actively manipulate qubits with precise quantum state control including adapted time management. The main challenges for a monolithic solution to qubit applications are the polarization manipulation of photons and the implementation of a variable on-chip time delay.}

\begin{figure*}[bth]
\begin{center}
\includegraphics*[width=0.85\textwidth]{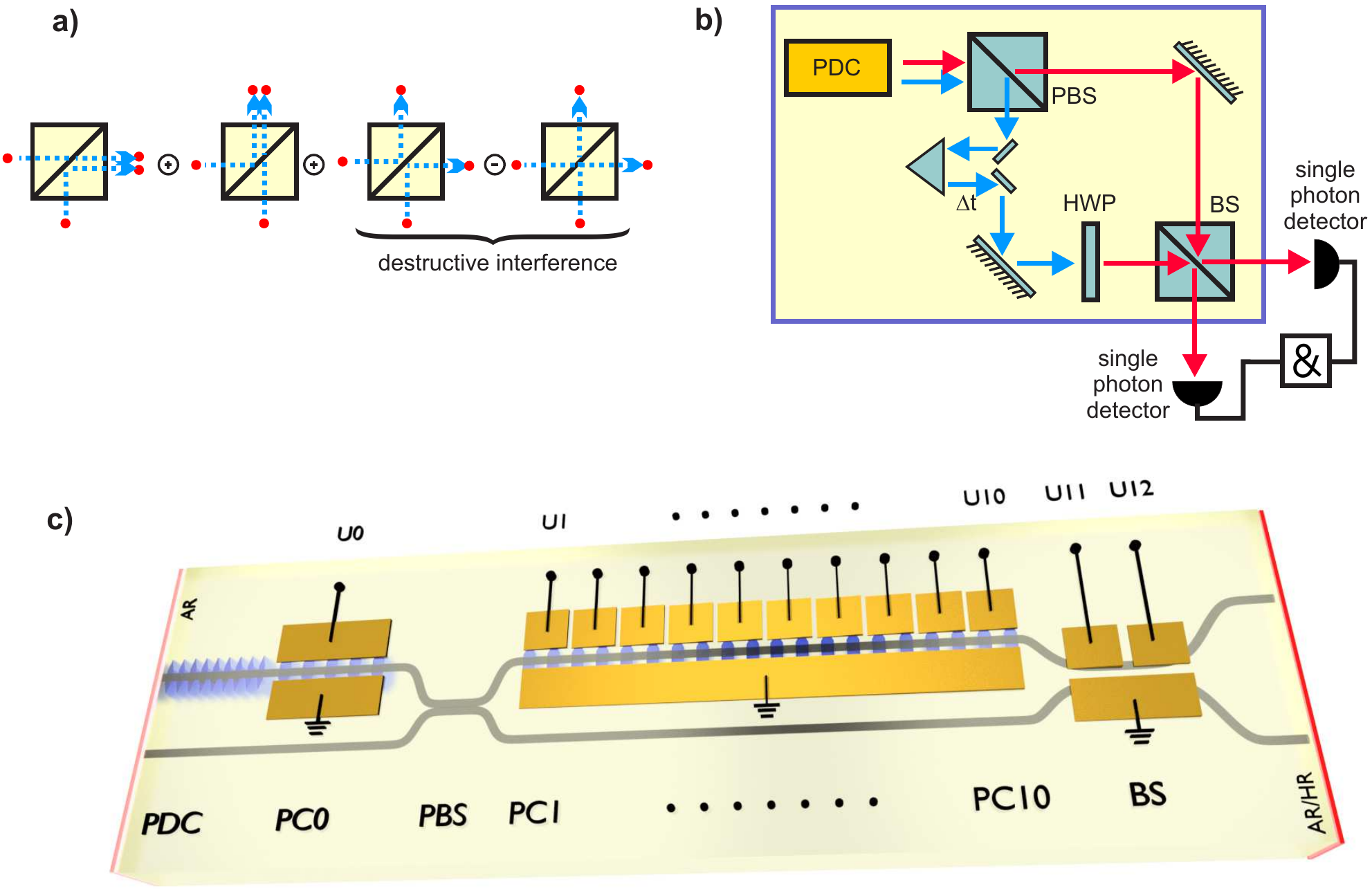}
\end{center}
\caption{
\label{HOM}\baselineskip12pt
\textbf{A miniaturized compact quantum circuit with active and accurate manipulation in \LN waveguides.}
\textbf{(A)}, HOM bunching effect of indistinguishable photons in a BS. \textbf{(B)}, Schematics of a typical HOM experiment using bulk optic components. Photon pairs with orthogonal polarizations and degenerate frequencies are generated via type II phase-matched PDC. A PBS is used to spatially separate the photons. To produce two identical photons, one photon is delayed in time, and its polarization is rotated by a half-wave plate (HWP) with respect to the other one. The photons are recombined again on a BS. All the functionalities in the yellow box are integrated into the chip. \textbf{(C)}, Scheme of the integrated quantum optical chip with monolithically integrated PDC source, electro-optic PCs, PBS, and \Mod{BS}. The gray lines denote the Ti-indiffused waveguides. In the periodically poled PDC section, orthogonally polarized photon pairs (H and V) are generated. In the subsequent PC$_0$, the complete conversion changes the polarization state of both photons from H to V and vice versa \Mod{via applying the control voltages U0}. These photons are spatially separated by the PBS. The H-polarized photons leave the splitter at the bar-state output, and V photons leave the splitter at the cross-state output. The H photons (at the bar state output of the splitter) enter into the segmented PC. At a certain position [depending on the voltages \Mod{(U1 to U10)} applied to the various segments] the polarization state is converted to V. Thus, these photons and the photons from the second branch enter the BS in V polarization. \Mod{BS is realized as a directional coupler with $\Delta\beta$ reversal electrodes. A 50:50 splitting ratio is precisely adjusted via the two control voltages U11 and U12. The waveguide end-faces are with \leftline{antireflection (AR) and high reflection (HR) coatings.}}
 }
\end{figure*}

As a comprehensive example to demonstrate deliberate manipulation of photonic states by precise polarization rotation and time regulation with all \Mod{necessary} properties in \Mod{a} single device, we will focus on Hong-Ou-Mandel (HOM) interference \cite{HOM1987}. As one of the most fundamental nonclassical experiments in quantum \Mod{optics, it} lies at the heart of many quantum logic operations \cite{WalmsleyS2017}, \Mod{for example}, boson sampling \cite{BroomeScience2013,SpringScience2013}; Bell-state measurement for quantum repeaters\cite{AzumaNC2015}; and \Mod{the Knill, Laflamme, and Milburn} protocol for quantum computing \cite{KLM2001}. However, an integrated chip comprising all functionalities needed for \Mod{manipulating quantum states on demand in} the complete HOM experiment has not yet been realized \Mod{so far}. In this work, we present an integrated electro-optic circuit \Mod{that can realize operations}, including photon pair state generation, passive routing, fast active polarization rotation for qubit manipulation, electro-optic balanced switching, and variable time delay management, on a single Ti:\LN waveguide chip. \Mod{Such fast electro-optically controllable on-chip time delays are crucial for all quantum applications, since temporal synchronization of manipulated state is a fundamental demand for all quantum logic operations.}

\section*{Results}
The HOM effect is well known as a nonclassical interference \Mod{phenomenon produced by} a beam splitter \Mod{(BS)}. When two identical single photons enter a 50:50 \Mod{BS} from opposite input ports, they bunch together and leave at the same output port (\Fig{HOM}~A). \Mod{In a familiar scheme to demonstrate this quantum effect in a bulk optical HOM experiment, photon pairs are generated} and then spatially separated via a polarization beam splitter (PBS), \Mod{as sketched in \Fig{HOM}~B}. After polarization rotation and a variable time delay \Mod{introduced} between the photons, they are recombined at a symmetric BS, where the quantum interference takes place.

Our \Mod{monolithic solution} is realized on the Ti:\LN platform \cite{LuoCLEO2018}, which exploits the strong $\chi^{(2)}$ nonlinearity for \Mod{both} photon pair generation and electro-optic manipulation of the qubits. To overcome an intrinsic birefringent delay in \Mod{the} nonlinear medium and \Mod{to} enable adjustable accurate \Mod{on-chip} time regulation, we introduce the concept of birefringent electro-optic delay (BED) that takes advantage of \Mod{the} electro-optic polarization conversion and the birefringence of the material itself.

\begin{figure*}[]
\begin{center}
\includegraphics*[width=0.85\textwidth]{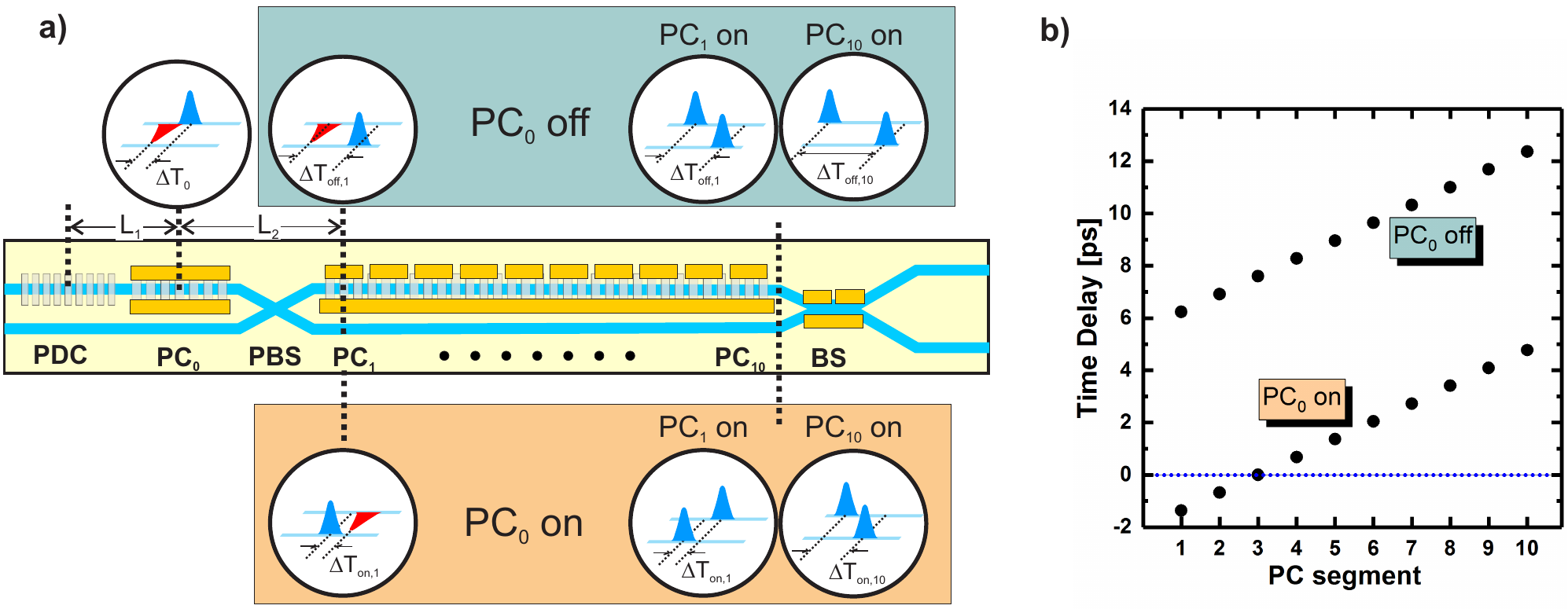}
\end{center}
\caption{\label{HOM-Timing} \baselineskip12pt \textbf{Illustration of the principle of the adjustable BED line.}
\textbf{(A)}, The diagram shows the chip design together with some insets illustrating the temporal relation of the horizontally (red) and vertically (blue) polarized photon wave packets at different positions of the structure and for various configurations of the PCs. Case I: If PC$_0$ is switched off, then the temporal walk-off increases along the structure. Thus, the time delay between the two photons can be varied depending \Mod{on} which element of the segmented converter is switched on; however, the two photons will never \Mod{arrive simultaneously at} the BS. Case II: If PC$_0$ is switched on, then the originally horizontally polarized photon can overtake the other photon before they arrive at the segmented PC. A simultaneous arrival of the two photons at the BS can be achieved if a certain element of the segmented PC is addressed. \textbf{(B)} Calculated time delay of the photons at the BS as \Mod{a} function of the element of the segmented PC, at which the final swapping of the polarization is performed. The diagram shows the result for the two cases \Mod{of} PC$_0$ on and off. The dotted line indicates the time synchronization between the two polarized photons. The parameters used for the calculations are adapted to the geometry of the fabricated device -- length of the PDC section (20.7 mm), PC$_0$ (7.62 mm), the PBS section (4.0 mm), and a single element of the segmented converters (2.54 mm). A group index difference $\Delta n_g=0.0805 $ has been derived \leftline{from the Sellmeier equations of \LN ($\lambda$=1551.7 nm).}}
\end{figure*}

\subsection*{HOM chip}
Our circuit design is shown \Mod{in detail} in \Fig{HOM}~C. \Mod{The} waveguides are fabricated by Ti indiffusion, which enables \Mod{single-mode} guiding in both polarizations. In a periodically poled waveguide section, spectrally degenerate photon pairs in the telecom range are generated using type II quasi-phase-matched \Mod{parametric down conversion} (PDC). In a specifically designed directional coupler which acts as \Mod{a} PBS, the orthogonally polarized photons are spatially separated. Because of the birefringence, the group velocities of the orthogonally polarized photons are different. Propagation along a waveguide of length $L$ results in a temporal walk-off $\Delta t=(n_{gH}-n_{gV})L/c$ with $n_{gH,V}$ being the group indices of the horizontally (H) and vertically (V) polarized photons, respectively. Usually, such a walk-off is unwanted in optical systems, \Mod{because temporal synchronization of qubits in a birefringent chip is required}. However, our device exploits exactly this walk-off for the adjustable timing management via BED, as detailed below. At the end of the \Mod{circuit}, the photons are routed to an electro-optic switch based on a directional coupler with $\Delta\beta$-reversal electrodes \cite{Kogelnik1976}, which acts as a balanced BS. Thus, trimming \Mod{of} the optimum 50:50 coupling can be achieved via the electro-optic control voltages.

\subsection*{Birefringent electro-optic delay}
The key elements of the BED system are \Mod{the} electro-optic polarization converters (PCs). Such a converter consists of a periodically poled waveguide with electrodes on each side \cite{HuangOE2007}. Via the nondiagonal  $r_{51}$ element of the electro-optic tensor, an electric field induces a periodic coupling of \Mod{the} orthogonally polarized modes and thus a wavelength-selective polarization rotation. Choosing the correct poling period enables quasi-phase matching, which compensates \Mod{for} the wave vector mismatch between the ordinary and extraordinary waves. Such a PC can be understood as the integrated optical implementation of a conventional folded \v{S}olc filter.

\begin{figure*}[htb]
\begin{center}
\includegraphics*[width=0.9\textwidth]{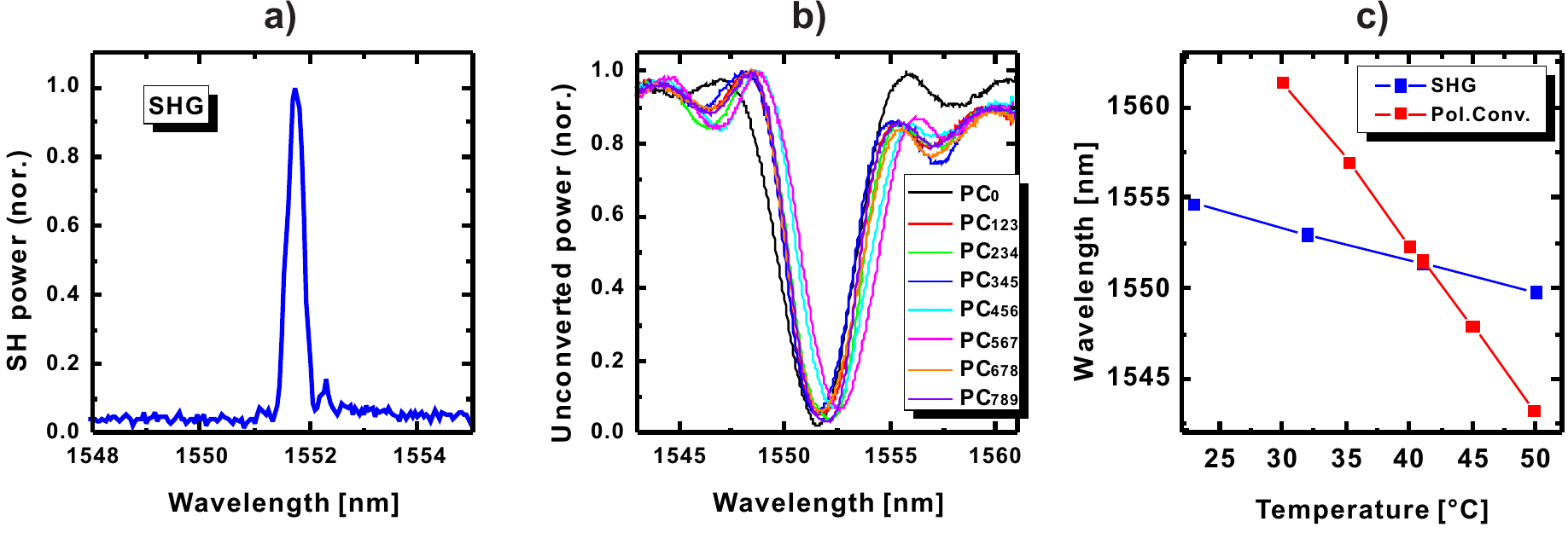}
\end{center}
\caption{\label{Classical} \baselineskip12pt \textbf{Classical characterization of the integrated circuit.}
\textbf{(A)}, Normalized power of  the second harmonic (SH) wave generated in the PDC section with a poling
period of $\Lambda_{PDC}=9.04~ \mu m$ as \Mod{a} function of the fundamental wavelength, which is from a tunable telecom laser with narrow bandwidth. \textbf{(B)} Spectral transmission characteristics of PC$_0$ and the various triple combinations of the segmented PC (with a poling
period of $\Lambda_{PC}=21.4~ \mu m$). We obtained the curves by launching \Mod{broadband incoherent light} in the telecom range and measuring the unconverted power behind a polarizer. The curves are normalized to a reference transmission spectrum obtained without conversion. \textbf{(C)}, Temperature dependence of the two phase-matching processes (PDC and PC). The crossing of the two curves \leftline{determines the optimum operation point, which is at $T = 43.6$~\degC\  and $\lambda$ = 1551.7 nm.}}
\end{figure*}

In our circuit, we \Mod{place the} first PC (PC$_0$) directly behind the PDC section, \Mod{followed by} a segmented PC comprising 10 elements (PC$_1$...PC$_{10}$) in one branch after the PBS (see \Fig{HOM}~cC). The \Mod{principle} of the adjustable BED system is illustrated in \Fig{HOM-Timing}~A. The group index difference $\Delta n_g$ causes a temporal walk-off between the photons. From the creation of the photon pairs in the PDC section to the center of PC$_0$, a walk-off of $\Delta T_1=\Delta n_g L_1/c$ arises. If PC$_0$ is off and the polarization state remains unchanged, then the walk-off increases monotonically with the propagation length, and the \Mod{photons in each pair never coincide} on the chip. By contrast, if PC$_0$ is on, then the polarization states get swapped. Then, the originally faster V-polarized photon propagates slower than an H-polarized photon and can be overtaken by the originally H-polarized (now V-polarized) photon. After PC$_0$, the H- and V-polarized photons are spatially separated at the PBS\Mod{\cite{SansoniQI2017}}. The H-polarized photons are routed to the branch with the segmented PC, which consists of a sequence of PC elements. In one of these elements, the polarization has to be flipped to ensure that both photons enter the subsequent BS in the same polarization (V polarization). Depending on which segment is chosen for the polarization conversion, the relative time delay between the two photons at the input ports of the BS can be finely adjusted.

\begin{figure*}[]
\begin{center}
\includegraphics*[width=0.85\textwidth]{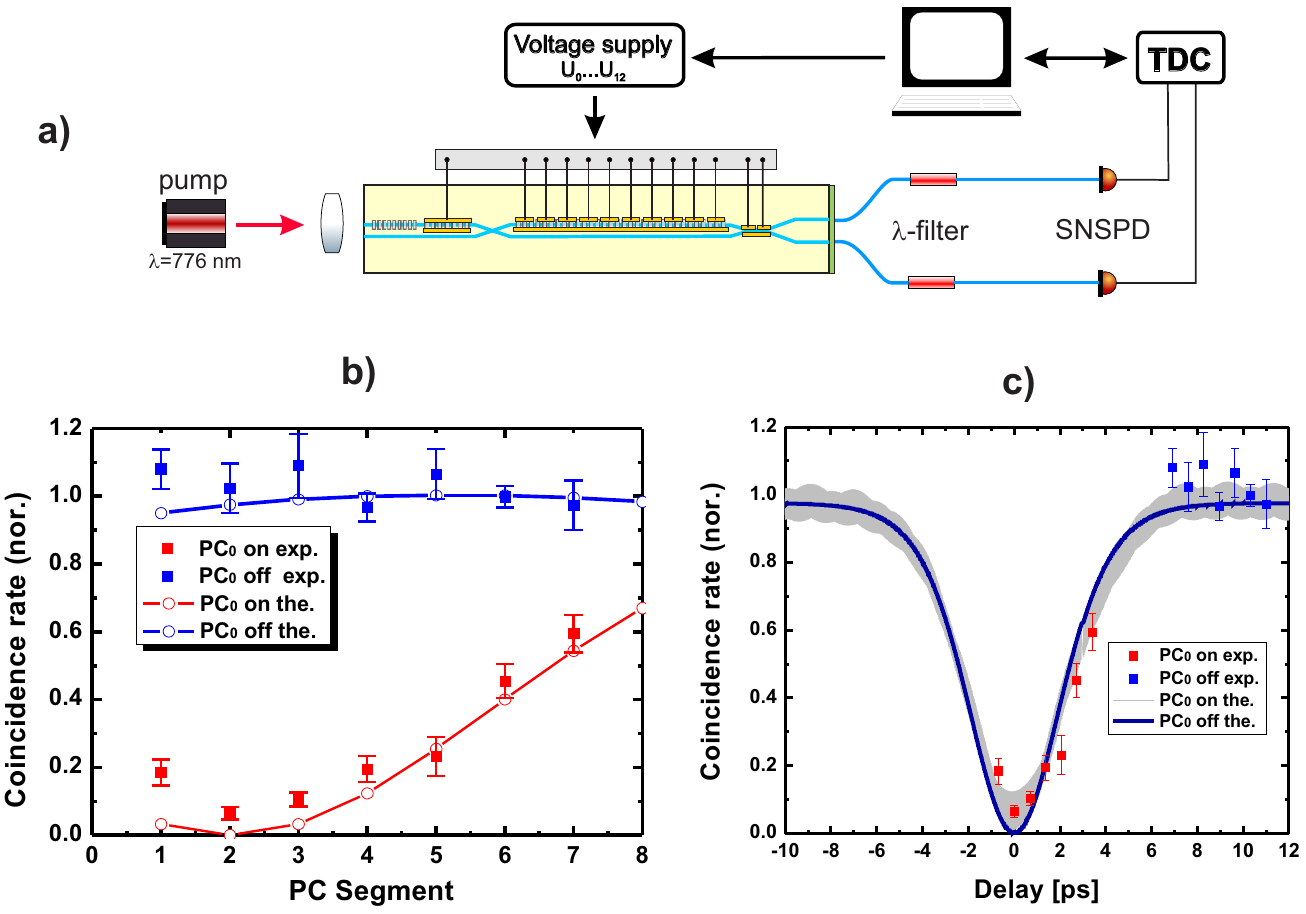}
\end{center}
\caption{\label{HOMdip} \baselineskip12pt \textbf{Experimental setup and quantum results.}
\textbf{(A)}, Experimental setup for quantum characterization of the active HOM chip. A tunable narrowband continuous-wave pump laser around 776 nm \Mod{is} coupled into the channel with the PDC source. To avoid higher order photon pair generation, the pump power \Mod{is} kept in the range of 100 $~\mu$W.
A temperature controller controls and stabilizes the previously determined temperature distribution of the sample. The two output ports from the chip \Mod{are} directly coupled \Mod{into} a pair of single-mode fibers. Via fiber-optical isolators to suppress the residual pump light and a 1.2 nm bandpass filter to suppress background photons, the transmitted photons are detected with superconducting nanowire detectors (SNSPDs) and time-to-digital converter (TDC). \textbf{(B)} Experimental and simulated results of the normalized coincidence rate as \Mod{a} function of which triple of the segmented PC is driven. The blue data and curve are for PC$_0$ off, while the red data and curve are for PC$_0$ on. \Mod{In the experiment, only seven triples of the segmented PC could be addressed because the electrode of PC10 was broken. Therefore, only 14 different delays were possible.} \textbf{(c)}, \Mod{Experimental and simulated profiles} of the HOM dip derived from the \leftline{coincidence results shown in (B) and the corresponding calculated time delay.}}
\end{figure*}

With the configuration of the present device, we can vary the \Mod{relative} time delay between $\sim$1.3~ps and more than 12~ps as shown in \Fig{HOM-Timing}~B. Although this variation is only possible in discrete steps, we can shift across the HOM dip with a sufficient resolution \Mod{down} to $\sim$0.6~ps.
The segmented PC consists of 10 electro-optic segments, which can be individually addressed; the length of each segment amounts to one-third of the length of the first PC (PC$_0$).  Thus, we obtain the same spectral characteristics as PC$_0$ if three subsequent segments (``triple") are driven simultaneously. Sliding such a triple across the 10 segments enables sampling at eight different delays; i.e., together with on/off switching of PC$_0$, we can, in total, set 16 different delays. The lengths of the different components are chosen such that they provide almost perfect temporal compensation at the BS when PC$_0$ is switched on and the second triple of the segmented PC (centered on PC$_3$) is driven.

\subsection*{Experimental characterization}

\Mod{Because a successful chip operation requires several phase-matched processes (the PDC and every PC), we need to characterize all the individual components separately with classical light to determine the optimum operating condition for the final experiments. First, we studied the phase-matching behavior of the PDC process. We used second harmonic generation (SHG), which is the reverse process to degenerate PDC, to identify the phase-matched wavelength (\Fig{Classical}~A) and its temperature dependence (\Fig{Classical}~C). To characterize the phase-matching conditions and to determine the required driving voltages of the various PC combinations, we used a bright broadband source to investigate the spectral transmission of the unconverted light (\Fig{Classical}~B) and its temperature dependence (\Fig{Classical}~C). The spectral width for PC$_0$ and all triples of the segmented PC is about $\sim$3.2~nm, which is substantially broader than the PDC spectrum ($\sim$1.3~nm). Although the central wavelengths are slightly different, which we attribute to inhomogeneities in the fabrication process, all conversion bands overlap reasonably well. From the temperature dependencies of the SHG process and the PC phase matching shown in \Fig{Classical}~C, we determined the operating point for the quantum experiments, i.e., chip temperature, pump wavelength, and driving voltages.}

\Mod{The entire measurement setup of the quantum experiment including external pump, fiber filters, and detection units is depicted in \Fig{HOMdip}~A. After suppressing the strong pump beam and reducing the background by the use of fiber filters, we recorded the
coincidence count rates between the two waveguide outputs.
The coincidence rate of the detectors should drop to zero when the two photons are perfectly identical in all properties. To prove the quantumness of the interference, the drop (dip) of the coincidences between perfectly identical and completely distinguishable cases should have a visibility beyond the classically expected value of 50\%. To quantify the measurement results without suffering from remaining uncertainties of the measurement conditions (such as power fluctuations, voltage drifting, temperature variations,etc.) we used the coincidence results for the completely distinguishable cases (i.e., longest time delays when $PC_0$ is off) as a reference and defined unit probability by averaging over many measurements.} \ModTo{Further details on the data evaluation are given in the Supplementary Materials. }

In \Fig{HOMdip}~B, we \Mod{plot the normalized experimental and simulated coincidence count rates} between the two waveguide outputs. Each point corresponds to sliding the triple from one segment to the next combination of the segmented PC. If the first PC (PC$_0$) is switched on (red data), we observe the predicted HOM dip \Mod{with a minimum coincidence count when the second segmented PC combination is switched on}. For comparison, we performed the same measurement with PC$_0$ switched off (blue data). In this case, we cannot achieve temporal synchronization, and we did not observe \Mod{any significant} variation of the count rate when sliding the triple. The normalized probability when PC$_0$ is off is constant \Mod{and close to the value of unit probability}. This means that, as expected, because of the further time delay between the arrival time of two photons at the BS, no HOM interference takes place.

In \Fig{HOMdip}~C, we \Mod{replot} the measured data and theoretical simulation as \Mod{a} function of the calculated time delay for the various segmented PC combinations. \Mod{The normalized coincidence rates of the first several segmented PC combinations (when PC$_0$ is off) are slightly larger than one, which could be attributed to partial indistinguishability together with spectral imperfections in the measurement. Because of the limitation of discrete variable time delay on-chip, we cannot resolve the real shape of the HOM dip in the experiment. With careful alignment of all control parameters, the visibility of the HOM interference, which is calculated between the lowest coincidence rate and unit probability, was 93.5 $\pm$ 1.8\% at a pump power of $\sim$100~$\mu$W. This value is significantly higher than the 50\% classical limit, evidencing the quantum nature of on-chip interference. Thus, although all the imperfections from each component together lead to a reduction of the visibility of the HOM dip, the two-photon interference is within the quantum regime.}

\Mod{Since the range of accessible timings does not allow us to measure the complete HOM dip, we double-checked the measurements by performing a careful simulation of the expected coincidence rates. To obtain reliable predictions, we implement a simulation based on unitary transformation of the initial wave function, taking into account the individual optical elements \cite{SharapovaNJP2017}. Actually, it is difficult to predict the exact HOM profiles with all realistic parameters taken into account. Apart from the three spectrally phase-matched properties described above, the individual imperfections of each optical element, such as the conversion efficiency of each PC, and the splitting ratio of PBS and BS are all included in our simulations. For simplicity, we have ignored the slight spectral shift of different PCs (as shown in \Fig{Classical}~B) and the consequent spectral mismatch among the PDC photons and fiber filters. From the simulations, we find that, in particular, the spectral change and a possibly incomplete conversion of PC$_0$ strongly affect the HOM dip. To account for an incomplete conversion at PC$_0$, we simulated the situations where PC$_0$ is on or off separately. Note that the imperfection of PC$_0$ could result in fast oscillations from interference between two components with the same polarization, which definitely cannot be resolved experimentally with our current discrete time delay resolution. However, the theoretically predicted shapes as a function of the time delay shown in \Fig{HOMdip}~C are in good agreement with our experimental data. The temporal width of the HOM dip is around 5.5~ps, which corresponds well to the predicted width for a 20.7~mm long PDC section after narrow filtering. Therefore, our programmable variable on-chip time delay of up to $\sim$12 ps enables us to synchronize two quantum states on-chip and to scan over half of the HOM dip width while preserving the quantum interference.}

\section*{Discussion}
We have demonstrated \Mod{a} quantum electro-optic circuit with active manipulation and adjustable time management of \Mod{the} photonic states, as \Mod{would be} needed in a monolithically  integrated structure. As an archetypical example, our two-photon HOM chip -- comprising a photon pair source, an active polarization manipulation, a programmable BED line, and a voltage-controllable BS -- completely and precisely recreates and surpasses the features of typical bulk implementations. Our work exhibits the capabilities of
the lithium niobate-based toolbox \Mod{for} creating, manipulating, and studying quantum states, tools for characterizing photonic states, and temporal encoding \Mod{of} the desired quantum states with precise timing. \Mod{In linear optical quantum computing, photon synchronization (particularly after propagation in many long delay lines and fast switching) requires fine tuning of the delay, which not only compensates for the mismatched lengths inside the circuits but also preserves quantum properties.} The successful realization of reconfigurable integrated electro-optic circuits opens \Mod{a door to the harnessing of} the tremendous potential of qubit manipulation in the \LN platform for future quantum science and technology. \Mod{In particular, even} when the PC is driven to yield an incomplete conversion, it can produce programmable superpositions of quantum states (i.e., the generic polarization-encoded qubit $\alpha \left| {\rm{H}} \right\rangle {\rm{ + }}\beta \left| {\rm{V}} \right\rangle$) for subsequent quantum logic operations \cite{CrespiNC2011}. \Mod{For example, when PC$_0$ only converts half of the polarization, hyperentanglement can be generated if the design parameters are carefully chosen \cite{SharapovaPRA2017}.} Moreover, because our device exploits the electro-optic effect in \LN, it also paves the way toward ultrafast processing which is well-established and widely acknowledged in classical communication devices such as modulators \cite{WootenIEEE2000} but rarely used in quantum optics \cite{BonneauPRL2012}.

\section*{Materials and methods}

\subsection*{Components and circuit design}
The integrated circuit is composed of several different components. All these components have already been demonstrated and optimized as individual devices. The integration into a complex circuit, however, \Mod{requires} more than just stacking the individual components together. Most significantly, the relative delay between the signal and idler photons must be adjustable with the segmented polarization controller. The accessible delay range must be large enough to cover at least half the width of the HOM dip. A second important criterion is the length of the overall structure, which should be kept as short as possible. The shorter it is, the easier it is to fabricate homogeneous structures.

Photon pairs were generated in the PDC section, which consists of a Ti-indiffused single-mode waveguide periodically poled with a period $\Lambda_{PDC}=9.04~ \mu m$ for quasi-phase matching. The type II phase-matched generation process provides orthogonally polarized photon pairs. The spectral bandwidth scales inversely with the device length. For the HOM chip, it is essential that the generated photon pairs are degenerate.  For a fixed poling period, the degeneracy point, i.e., the wavelength at which degenerate PDC is phase matched, can be tuned by varying the temperature. The tuning slope is about -0.15~nm/\degC\ (see \Fig{Classical}).

An integrated version of a PBS can be realized by a directional coupler. For the HOM chip, we used a ``zero-gap" coupler, i.e., a structure in which the two incoming waveguides merge into a broader single waveguide in the coupling section before they separate again. By selecting a proper length of the coupling section and taking \Mod{the} coupling in the branching ranges into account, we \Mod{can} design a compact structure that provides routing of the H-polarized photons to the cross-state output of the coupler, whereas the V-polarized photons are directed into the bar-state output.

For the PC the electro-optic properties of \LN are exploited. The structure of a PC consists of a periodically poled waveguide with a poling period $\Lambda_{PC}=21.4~ \mu m$. Adjacent to
both sides of the waveguides are electrodes deposited on the sample. Applying a voltage $U$ to these electrodes
can induce a polarization conversion; however, this process requires phase-matching as well. The corresponding phase-matching condition is given by $\beta_V-\beta_H\approx2\pi/\Lambda_{PC}$ with $\beta_{V,H}$ being the propagation constants of the V- and H-polarized photons, respectively. This phase-matching requirement makes the PC wavelength dependent. Again, the spectral bandwidth is \Mod{inversely proportional} to the device length. The temperature dependence of the phase matching, which can be exploited to spectrally tune the device, is much stronger than that of the PDC phase matching. The slope is about -0.7~nm/\degC.

The efficiency of the conversion is determined by the overlap of the optical mode fields with the electric field, the electric field strength, and the length of the device. Thus, the efficiency can be adjusted by the applied voltage. For our PCs we typically get a voltage-length product of about 15~V$/$cm, i.e., a drive voltage of 15~V is sufficient to obtain complete polarization conversion in an 1-cm-long device.

For the BS, we used another type of directional coupler with a 6,000-$\mu$m-long coupling section composed of two waveguides separated by a 6-$\mu$m-wide gap. With a split electrode on top of the waveguide, this $\Delta\beta$-reversal structure enables us to adjust the splitting ratio via two control voltages.

For the overall circuit design, the spectral bandwidth of the PDC process and the PC must be \Mod{matched}. The PDC bandwidth must be smaller than the PC bandwidth to make sure that the PDC states are fully converted. We used a 20,700-$\mu$m-long PDC section from which the expected PDC bandwidth is around 1.3 nm. The length of PC$_0$ and of each of the triples in the segmented PC is 7620~$\mu$m corresponding to a conversion bandwidth of 3.2 nm. The length of the PBS section is  4000~$\mu$m. The BS together with the branching of the output waveguides to a separation of 127~$\mu$m has an overall length of 13,100~$\mu$m. Thus, the length of the overall structure is slightly more than 70~mm.

\subsection*{Sample fabrication}
First, the optical waveguides were fabricated. Low-loss single-mode waveguides in both polarizations were obtained by indiffusing 7-$\mu$m-wide, 80-nm-thick titanium stripes into the z-cut \LN substrate at 1060~\degC\ for 9 hours. Subsequently, the periodic poling of the PDC and the PC sections was performed by field-assisted ferroelectric domain inversion. Afterward, a 400-nm-thick planar SiO$_2$ buffer layer was deposited on top of the sample, which prevents excess losses of the optical waves due to the metal electrodes of the electro-optic components. A 200-nm-thick planar aluminium layer deposited on top of the buffer layer was lithographically patterned to form the various electrodes. After polishing the waveguide end-faces, we deposited an antireflection coating for the pump wave \Mod{on} the input side, and \Mod{on} the output side a dielectric multistack coating providing antireflection for the 1550-nm range and high reflectivity at the pump wavelength for pump suppression. The overall chip was assembled in a temperature-stabilized mount, and the electrodes were contacted via wire bonding.

\subsection*{Chip characterization and experimental details}
The waveguide loss is around 0.1~dB/cm for both TE and TM polarized light. The splitting ratio of the PBS is larger than 17~dB, and the conversion ratio of PC is typically around 20~dB. The pump suppression can reach more than 100~dB by using the combination of waveguide end-face coating, fiber-optical isolators, and a bandpass fiber filter. From the waveguide outputs to the detectors, the loss is in the range of 7~dB, which includes 2~dB from fiber butt coupling, 3~dB from cascaded isolators and filters, and another 1.5~dB from the detection systems. Together with the individual excess losses of PBS and BS, which are typically below 0.5~dB, the total loss is around 11~dB, which corresponds to the Klyshko efficiency of 5\% obtained from coincidence measurements. At a pump power of 100~$\mu$W, the typical single count rate is about 2~kHz, while the coincidence rate is in the range of 100~Hz. From the coincidence measurements, the brightness of the PDC source is estimated to be around $3\times10^5$~pairs/(s$ \cdot $mW$ \cdot $nm).

\subsection*{Theoretical simulation}
More details about the analysis of theoretical simulation are shown in the Supplementary Materials.

\section*{Supplementary Materials}

\begin{itemize}
  \item section S1. Theoretical simulation
  \item section S2. Dispersion properties of electro-optic PCs
  \item section S3. Measurement procedure and normalization
  \item fig. S1. Simulated profiles of the HOM-dip.
  \item fig. S2. Raw single rates and coincidence rates.
\end{itemize}

\subsection*{section S1: Theoretical simulation}

For theoretical simulations of the HOM dip we consider the unitary evolution of the PDC state created in the periodically poled section of the \LN waveguide. The unitary evolution contains the PBS, BS transformations, free propagation inside the waveguides and frequency dependent transformation at all PCs. After passing all optical elements the final state is projected into the detectors using  positive-operator valued measures. The coincidence probability of two simultaneous detections is calculated versus time delay. The time delay is created by switching one by one the set of segmented PCs. In the ideal case, when all optical elements work perfectly and the PCs are frequency independent, the HOM dip drops to zero and has a smooth form.  However, the total interference profile is very sensitive to realistic imperfections and experimental conditions, in particular to the first PC (PC$_0$) which causes not only a reduction of the visibility but also leads to a more complex interference pattern.

\Mod{
The profile of the HOM-dip depends on the presence of filters and imperfections of the elements inside the circuit, as shown in Fig~S1. If all elements are ideal and there is no filter in the system, the HOM-dip has a triangular shape. This is due to the two-photon spectral distribution of the PDC photon which is characterized by the sinc-shaped phase-matching function. By using different shapes of filters, it is possible to modify the profile of the HOM-dip, since the original spectral distribution of PDC photons is cut and reshaped. For example, if we use a rectangular filter to cut the lobes of the sinc-function, the triangular shape of the HOM-dip becomes more sinc-like. It should be noted, however, that such filtering could result in a coincidence probability larger than 0.5 for certain time delays.}

\begin{figure}[htb]
\begin{center}
\includegraphics[width=0.8\columnwidth]{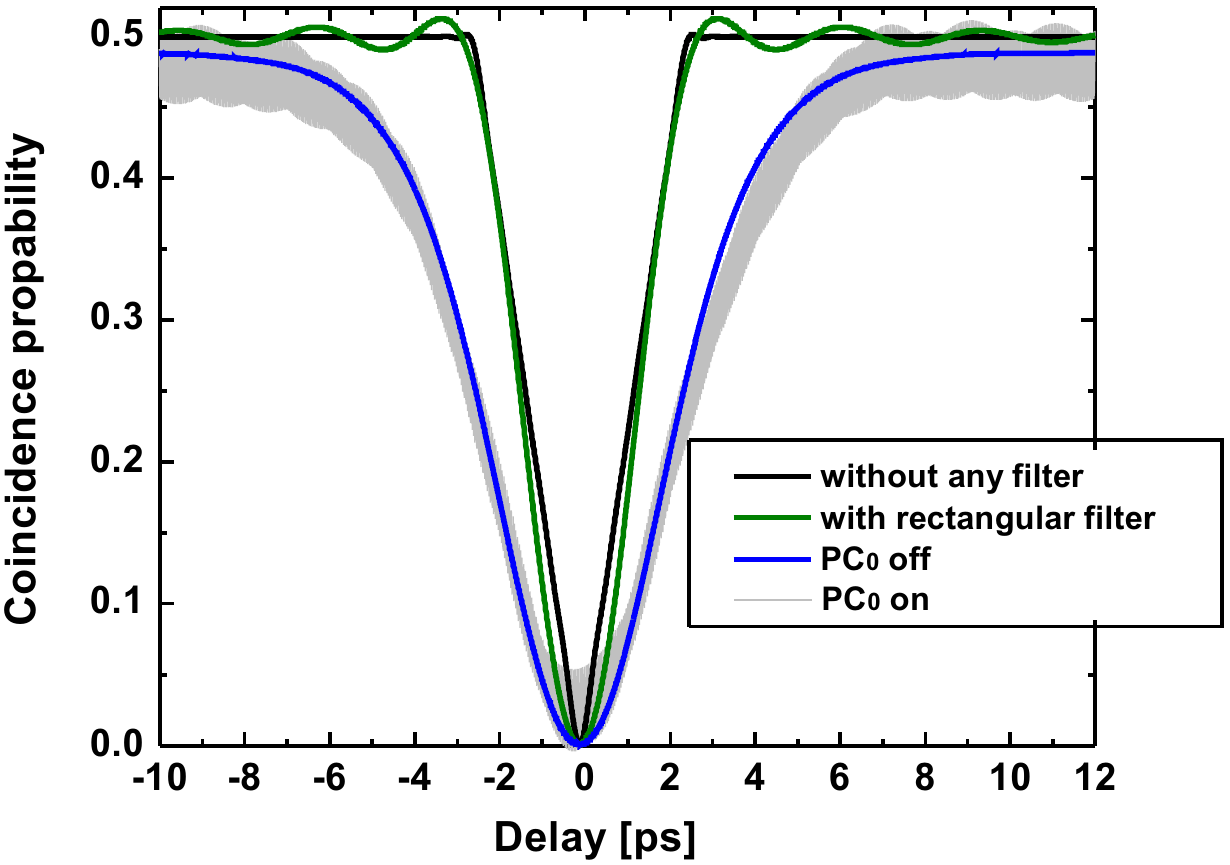}
\end{center}
\captionsetup{labelformat=empty}
\caption*{Figure S1: \Mod{\bf{Simulated profiles of the HOM-dip}}. \Mod{The diagram shows calculated HOM-dips, i.e.\ the coincidence probability versus the relative time delay, for an ideal sinc-shaped PDC source with 1.3 nm bandwidth without any spectral filters (black), with an additional rectangular 2.3 nm wide spectral filter (green), with sinc-profile filtering using a segmented PC and an additional 1.2 nm wide Lorentz fiber filter when PC$_0$ is off (blue), and sinc-profile filtering \leftline{with two PCs and the fiber filter when PC$_0$ is on (grey).}}}
\label{fig:suppl}
\end{figure}

\Mod{Not only does filters change the shape of the HOM-dip, but a realistic PC does so as well.
It is important to note that since the PC depends on the phase-matching condition, the conversion is intrinsically frequency dependent, which results in conversion occurring only in a narrow wavelength range.
When PC$_0$ is off, the PDC spectrum is modified by the filtering caused frequency
dependent segmented PC and a Lorentz fiber filter with bandwidth of 1.2 nm. The filtered narrower profile is very close to a Gaussian, as reflected in the broader Gaussian shape of the HOM dip. Also, in this case the constant value of the HOM-dip is less than 0.5 because of losses in the conversion process (in comparison with an ideal PC). We normalize the curve at 0.5 for comparison with the ideal shape. When PC$_0$ is on, the incomplete conversion of the photons of both polarizations plus the spectral reshaping by the combination of all filters makes the HOM-dip become broader and fatter modulated with fast oscillations.}

\ModTo{
\subsection*{section S2: Dispersion properties of electro-optic PCs}
To understand the dispersive effects related to the electro-optic PC we can look into the solution of the coupled mode equations. The dispersive properties are usually specified by the dependence of spectral phase $\Phi$ as function of the wavelength, or more common, as function of the angular frequency. For this purpose, this phase dependence can be expanded into a Taylor series according to:}
\setcounter{equation}{0}
\renewcommand{\theequation}{S.\arabic{equation}}
\begin{eqnarray}
\Phi {\rm{ = }}{\Phi _0}\left( {{\omega _0}} \right){\rm{ + }}\mathop {\left. {{{\partial \Phi } \over {\partial \omega }}} \right|}\nolimits_{{\omega _0}} \left( {\omega  - {\omega _0}} \right) + {1 \over 2}\mathop {\left. {{{{\partial ^2}\Phi } \over {\partial {\omega ^2}}}} \right|}\nolimits_{{\omega _0}} {\left( {\omega  - {\omega _0}} \right)^2} +  \cdots.
\label{eq:PhiTaylor}
\end{eqnarray}
\ModTo{The constant term ${\Phi _0}\left( {{\omega _0}} \right)$ is just an overall phase-shift. The linear one only describes the time delay and has no impact on the shape of the pulse envelope. The quadratic term is the group
delay dispersion (GDD), which is usually quantified by the second derivative of $\Phi$ and expressed in units fs$^2$.}

\ModTo{Applying these definitions for quantifying the dispersive properties of the electro-optic converter, we can easily calculate the corresponding derivative taking the solutions of the coupled mode theory. We obtain}
\begin{eqnarray}
{{\partial \Phi } \over {\partial \omega }}{\rm{ = }}{L \over {2c}}\left( {{n_{g,TE}} - {n_{g,TM}}} \right),
\label{eq:1stTaylor}
\end{eqnarray}
\ModTo{with ${n_{g,i}},$ being the group index of the respective polarization. In a similar way, we can determine the GDD by calculating the second derivative}
\begin{eqnarray}
\rm{GDD} = {{{\partial ^2}\Phi } \over {\partial {\omega ^2}}}{\rm{ = }} - {{{\lambda ^2}L} \over {4\pi {c^2}}}\left( {{{\partial {n_{g,TE}}} \over {\partial \lambda }} - {{\partial {n_{g,TM}}} \over {\partial \lambda }}} \right),
\label{eq:GDD}
\end{eqnarray}
\ModTo{To estimate numbers we can use Eq.~\ref{eq:GDD} and insert realistic parameters. For $L = 7.5$~mm and $\lambda = 1550$~nm the calculated GDD is about $-44$~fs$^2$, which is completely neglectable in our experiment.}

\ModTo{
\subsection*{section S3: Measurement procedure and normalization}
The value of visibility depends on the normalization procedure which is used. To normalize the HOM-dip, we had to determine a reference value. For this purpose, we first measured the coincidence rate ($\sim$ 100~Hz) for the longest possible time delay, i.e. with PC$_{0}$ off and PC$_{789}$ on. To ensure that the time delay is long enough to be completely outside of the HOM dip, we repeated the same measurement with non-degenerate photons, i.e. we slightly detuned the PDC source. As both measurements yielded the same rates at the same pump power, we used them as reference to define unit probability. In Fig. 4 we scaled the measured data to this unit probability and we determined the HOM visibility from the probability at zero-time delay compared to unit probability.}

\ModTo{
In Fig. S2 an example of a recorded raw data set is shown. The left diagram shows the single count rates at the detectors D$_{1}$ and D$_{2}$ as function of the driven PC segment for PC$_{0}$ on and off.  In the right diagram  the corresponding coincidence counts are shown.}

\begin{figure}[htb]
\begin{center}
\includegraphics[width=0.9\columnwidth]{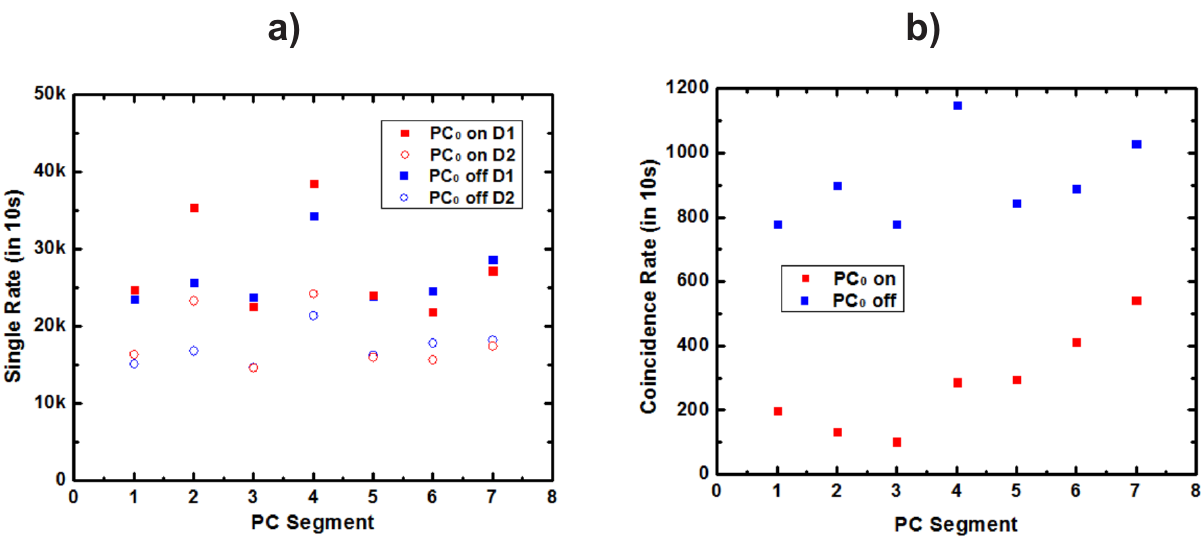}
\end{center}
\captionsetup{labelformat=empty}
\caption*{Figure S2: \ModTo{\bf{Raw single rates and coincidence rates}}. \ModTo{a) The raw single count rates from two detectors D$_1$ (solid square) and D$_2$ (hollow circle) in 10 seconds, when PC$_0$ is on (red) and off (blue). b) The raw coincidence rates \leftline{from two scenarios: PC$_0$ is on (red) and off (blue).}}}
\label{fig:suppl}
\end{figure}

\bibliography{scibib}

\bibliographystyle{Science}

\section*{}
\textbf{Acknowledgments:} We thank S. Krapick and R. Driben for helpful discussions and technical assistance, and L.-A. Wu, E. Meyer-Scott, J. M. Donohue, T. Bartley and B. Brecht for proofreading the manuscript. P.R.S. thanks the state of Nordrhein-Westfalen for support by the Landesprogramm f\"ur geschlechtergerechte Hochschulen. \textbf{Funding:} This work was funded by the Deutsche Forschungsgemeinschaft (DFG, German Research Foundation)
¨C Projektnummer 231447078 ¨C TRR 142 (via project C02) and the Gottfried Wilhelm Leibniz-Preis (grant SI1115/3-1). \textbf{Author contributions:} K.-H.L., S.B., and H.H. built the experiment and collected the data. C.E. and R.R. contributed to the fabrication of sample. P.R.S. and T.M. provided the theoretical support. H.H. and C.S. conceived the original idea. C.S. supervised the project. K.-H.L. and H.H. drafted the manuscript. All authors discussed and contributed to the final version of the manuscript. \textbf{Competing interests:} The authors declare that they have no competing interests. \textbf{Data and materials availability:} All data needed to evaluate the conclusions in the paper are present in the paper and/or the Supplementary Materials. Additional data related to this paper may be requested from the authors.

\end{document}